\documentclass[aps, twocolumn, prd, preprintnumbers, amsmath, amssymb, amsfonts, superscriptaddress, nofootinbib]{revtex4-1}
\pdfoutput=1
\usepackage{amsmath,amssymb,bm,mathrsfs}
\usepackage{hyperref}
\usepackage{url}
\usepackage{graphicx}
\usepackage{color}
\usepackage{tikz}
\usepackage[compat=1.1.0]{tikz-feynman}
\usepackage[normalem]{ulem}

\definecolor{Orange}{cmyk}{0,0.61,0.87,0}
\definecolor{JungleGreen}{cmyk}{0.99,0,0.52,0}
\definecolor{OliveGreen}{cmyk}{0.64,0,0.95,0.40}
\definecolor{Brown}{cmyk}{0,0.81,1,0.60}
\definecolor{RoyalBlue}{cmyk}{0.71,0.53,0,0.12}
\definecolor{Gray}{cmyk}{0,0,0,0.40}
\definecolor{LightPink}{cmyk}{0.0,0.25,0,0}
\definecolor{LLightPink}{cmyk}{0.0,0.10,0,0}
\definecolor{LightBlue}{cmyk}{0.25,0,0,0}
\definecolor{LightGray}{cmyk}{0,0,0,0.2}

\newcommand{\mr}[1]{\mathrm{#1}}
\newcommand{\mc}[1]{\mathcal{#1}}

\begin{document}
\hfill TU-1289, KEK-QUP-2025-0027

\title{
Gravitational Decays of Secluded Scalars and Graviton Dark Radiation
}

\author{
Kazunori Nakayama
}
\email{kazunori.nakayama.d3@tohoku.ac.jp}
\affiliation{Department of Physics, Tohoku University, 
Sendai, Miyagi 980-8578, Japan} 
\affiliation{International Center for Quantum-field Measurement Systems for Studies of the Universe and Particles (QUP), KEK, Tsukuba, Ibaraki 305-0801, Japan}

\author{
Fuminobu Takahashi
}
\email{fumi@tohoku.ac.jp}
\affiliation{Department of Physics, Tohoku University, 
Sendai, Miyagi 980-8578, Japan} 
\affiliation{Kavli IPMU (WPI), UTIAS, University of Tokyo, Kashiwa 277-8583, Japan}

\author{Juntaro Wada}
\email{juntaro.wada.e5@tohoku.ac.jp }
\affiliation{Department of Physics, Tohoku University, Sendai, Miyagi 980-8578, Japan}
\affiliation{Technical University of Munich (TUM), School of Natural Sciences, Physics Department, James-Franck-Str. 1, 85748 Garching, Germany,
}

\begin{abstract}
We discuss graviton dark radiation produced by the decay of a secluded scalar field that couples to the Standard Model (SM) only through gravity. Such scalar fields are long-lived, and their decay channels generically include gravitons. If such particles existed and dominated the early universe, a sizable branching ratio into gravitons would yield non-negligible dark radiation that significantly alters the subsequent thermal history of the universe. In this work, we focus on the dark glueball as a representative secluded hidden scalar and compare the decay rates into SM particles via a non-minimal coupling to gravity with those into gravitons, paying attention to how the breaking of conformal invariance affects the amount of graviton dark radiation. We find that decays into the SM are dominated by two-body decay channels into Higgs bosons and gluons. In particular, when the Higgs field has a large non-minimal coupling to gravity, the production of graviton dark radiation is naturally suppressed in the metric formalism, and the SM sector is preferentially reheated and energy transfer to other hidden sectors is suppressed. Finally, we present the expected gravitational-wave spectrum resulting from dark glueball domination. 
\end{abstract}

\maketitle

%%%%%%%%%%%%%%%%%%%%%%%
\section{Introduction}
%%%%%%%%%%%%%%%%%%%%%%%
The nature of the interactions between the Standard Model (SM) and possible hidden sectors remains one of the key open questions in particle cosmology. While many well-motivated extensions assume the presence of mediator particles carrying both SM and hidden charges, it is also plausible that the hidden sector communicates with the visible particles only through gravity~\cite{Ema:2015dka, Ema:2016hlw, Ema:2018ucl, Ema:2019yrd, Redi:2020ffc, Gross:2020zam, Garani:2021zrr, Clery:2021bwz, Clery:2022wib}.
Such a secluded sector is especially interesting, since its cosmological signatures originate solely from gravitational effects, providing a unique way to explore physics that cannot be tested in laboratory experiments.

A scalar field that couples to the SM solely via gravity is naturally long-lived, and its decay channels inevitably include gravitons, unless such operators are forbidden by symmetry~\cite{Ema:2021fdz}.  If such a field dominated the energy density of the universe at some epoch, its decay could inject a large amount of high frequency gravitons which contributes to the effective number of relativistic degrees of freedom, $\Delta N_{\mathrm{eff}}$, which is the important observable in the early universe~\cite{Smith:2006nka, Sendra:2012wh, Pagano:2015hma, Clarke:2020bil}.\footnote{See, e.g., Ref.~\cite{Ichikawa:2007jv} for cases where scalar fields decay into other light degrees of freedom such as axions, which can contribute to $\Delta N_{\rm eff}$.}  The resulting constraints from cosmic microwave background (CMB) and baryon acoustic oscillation (BAO) observations, such as those from Planck~\cite{Planck:2018vyg} and DESI~\cite{DESI:2024hhd}, therefore provide a sensitive probe of such scenarios.

Previous studies have explored various cosmological consequences of secluded scalar fields coupled only gravitationally~\cite{Redi:2020ffc, Gross:2020zam, Garani:2021zrr, Ema:2021fdz}.  In particular, it has been pointed out that their decay into gravitons can lead to potentially observable dark radiation and high-frequency stochastic gravitational waves~\cite{Ema:2021fdz, Tokareva:2023mrt, Mudrunka:2023wxy, Strumia:2025dfn}. 
%However, 
Importantly, the amount of graviton dark radiation crucially depends on how conformal symmetry is broken in both the hidden and visible sectors, which determines the competing decay channels into SM particles.\footnote{
    Other scalar-induced graviton production channels include graviton bremmstrahlung~\cite{Nakayama:2018ptw,Barman:2023ymn,Barman:2023rpg,Hu:2024awd}, pair annihilation into gravitons~\cite{Ema:2015dka,Ema:2020ggo,Choi:2024ilx}, scattering of decay products off the parent scalar~\cite{Xu:2024fjl,Bernal:2025lxp}. For our cases of interest, they are all subdominant compared to the scalar decay into the graviton pair.
}

In this work, we focus on the dark glueball as a representative example of a secluded scalar field.  We examine how its decay rates into the SM sector and into gravitons are controlled by the breaking of conformal invariance, especially through a non-minimal coupling of the Higgs field to gravity and the conformal (trace) anomaly.  We analyze both the metric and Palatini formulations and derive the corresponding decay widths induced by the trace anomaly and higher-curvature operators.  We then evaluate the resulting amount of graviton dark radiation and identify the parameter region consistent with the current limits on $\Delta N_{\mathrm{eff}}$.  Finally, we present the predicted spectrum of high-frequency gravitational waves originating from the decay of the dark glueball, which can provide a target for future gravitational-wave observations.

This paper is organized as follows.  In Sec.~\ref {sec: model}, we introduce the model setup, including the distinction between the metric and Palatini formulations.  In Sec.~\ref {sec: secluded scalar decay}, we compute the decay rates of the secluded scalar into the SM particles, emphasizing the role of conformal symmetry breaking.  The production of graviton dark radiation and the corresponding cosmological constraints are discussed in Sec.~\ref {sec: graviton as dark radiation}, followed by the analytical computation of the stochastic gravitational-wave spectrum in Sec.~\ref {sec: stochastic gravitational wave spectrum}.  We summarize our findings and discuss possible extensions in Sec.~\ref {sec: discussion and conclusions}.

%%%%%%%%%%%%%%%%%%%%%%%
\section{Model}
\label{sec: model}
%%%%%%%%%%%%%%%%%%%%%%%
In this section, we set up a simple framework for a scalar field $\phi$ that couples to the SM sector only through a non-minimal coupling to gravity.  Such a scalar may be an elementary singlet, or it may arise as a glueball in a secluded dark gauge sector.  Our setup is intended to describe a generic scalar of this type, and our analysis can be applied to any such scalar field.  When we estimate the size of the couplings induced by conformal symmetry breaking, both for the non-minimal coupling that appears below and for the higher-dimensional operator that generates the decay into a graviton pair, we will have the dark glueball case in mind and assume that they are controlled by the dynamical scale of the dark gauge sector.
We will refer to this scalar either as “the scalar field’’ or as “the dark glueball’’ interchangeably.

Concretely, when we have the dark glueball case in mind, we consider an extension of the SM by introducing a pure dark gauge sector, $G_{\mathrm{dark}}$.  Let $\Lambda$ denote the confinement scale of $G_{\mathrm{dark}}$.  Dark glueball models with purely gravitational interactions have been discussed in Refs.~\cite{Redi:2020ffc, Gross:2020zam, Garani:2021zrr}, but here we are interested in the effect of conformal symmetry breaking on the decay rate. Therefore, we assume that the Higgs field $H$ also has a non-minimal coupling to the Ricci scalar $\hat{R}$ in the Jordan frame, which was not considered in Refs.~\cite{Redi:2020ffc, Gross:2020zam, Garani:2021zrr}. Then, the action in this model can be written as follows:
\begin{align}
\label{eq: action non minimal-couple to gravity}
S &= \int d^4 x \sqrt{-\hat{g}} \left[{M_{\mr{pl}}^2 \over 2}\Omega^2 \hat{R} + \mc{L}_{D} + \mc{L}_{SM} \right],\\
\Omega^2 & := 1 - {2 \Lambda \phi \over M_{\mr{pl}}^2} - {2 \xi |H|^2 \over M_{\mr{pl}}^2},
\end{align}
where $\xi$ is the non-minimal coupling constant, $\hat{g}_{\mu\nu}$ denotes the metric in the Jordan frame, and $M_{\mr{pl}} := (8 \pi G)^{-1/2}$ is the reduced Planck mass. The non-minimal coupling $\xi$ is bounded as $|\xi| \lesssim 10^{15}$ by collider experiment~\cite{Atkins:2012yn, Xianyu:2013rya}. We also denote $\mc{L}_{D}$ and $\mc{L}_{SM}$ the Lagrangian for the dark sector and SM sectors without non-minimal coupling to gravity.

Among the terms in the SM Lagrangian, the relevant contributions to the dominant decays of $\phi$ come from the Higgs and gauge boson sectors, $\mc{L}_{\mr{Higgs}} + \mc{L}_{\mr{Gauge}}$. These are given by
\begin{align}
&\mc{L}_{\mr{Gauge}} 
    = - {1 \over 4} G_{\mu\nu}^a G^{a\,\mu\nu} 
    - {1 \over 4} W_{\mu\nu}^a W^{a\,\mu\nu}
     - {1 \over 4} B_{\mu\nu} B^{\mu\nu}, \\
&\mc{L}_{\mr{Higgs}}
    = - g^{\mu\nu} (D_{\mu} H)^{\dagger}  (D_{\nu} H) - V(H),
\end{align}
 where $G_{\mu\nu}^a, W_{\mu\nu}^a, B_{\mu\nu}$ are the field strengths of the gluon $G_{\mu}$, weak gauge boson $W_{\mu}$, and hypercharge gauge boson $B_{\mu}$, respectively.
The covariant derivative of the Higgs field is defined as
\begin{align}
D_{\mu} H := \left(\partial_{\mu} - i g W_{\mu}^a {\tau^a \over 2} - i {1 \over 2} g' B_{\mu} \right) H,
\end{align}
where $\tau^a/2$ are the generators of $SU(2)$ and $g$ and $g'$ denote $SU(2)_L$ weak and the $U(1)_Y$ hypercharge gauge coupling. $V(H)$ denotes the Higgs potential. 
In this work, we are interested in the case where the scalar field has a mass well above the Higgs vacuum expectation value (VEV), $m_{\phi} \gg v$. In this situation, the effects of conformal-symmetry breaking induced by the Higgs VEV are typically suppressed by $(v / m_{\phi})^2$ and can be safely neglected. Hereafter, we therefore neglect the Higgs VEV in our analysis.

\subsection{Metric formulation}
We denote $g_{\mu\nu}$ as the metric in the Einstein frame and perform a Weyl transformation,
\begin{align}
\label{eq: weyl transformation}
g_{\mu\nu} = \Omega^2 \hat{g}_{\mu\nu}.
\end{align}
In the metric formalism, the Ricci scalar transforms as
\begin{align}
\hat{R} = \Omega^{2} \left( R + 6 \Box \ln \Omega - 6 g^{\mu\nu} \partial_{\mu} \ln \Omega \, \partial_{\nu} \ln \Omega \right),
\end{align}
where $R$ is the Ricci scalar in the Einstein frame and $\Box := g^{\mu\nu}\nabla_{\mu}\nabla_{\nu}$ denotes the d’Alembertian operator with the covariant derivative $\nabla_{\mu}$.

After the Weyl transformation, (gravity-mediated) interaction between $\phi$ and the Higgs field appears in the Einstein frame.  
Expanding $\Omega$ in this action around $\phi=0,  H=0$ and flat spacetime, we obtain the following interaction term between the scalar field and the Higgs field:
\begin{align}
S \supset 
\label{eq: mixing between phi and Higgs}
    \int d^4 x {2 \Lambda \over M_{\mr{pl}}^2} 
    (3 \xi \Box \phi |H|^2 - \phi |\partial_{\mu} H|^2).
\end{align}

In terms of canonically normalized Higgs field by $H_c:= \Omega^{-1} H$,  this interaction is rewritten as
\begin{align}
\label{eq: mixing between phi and canonical Higgs}
S \supset 
    \int d^4 x {\Lambda \over M_{\mr{pl}}^2} 
    (6 \xi -1)\left(\Box \phi \right) |H_c|^2.
\end{align}

Next, concerning the gauge sector, in four-dimensional spacetime, there is no tree-level coupling between $\phi$ and the gauge bosons. However, as pointed out in Refs.~\cite{Endo:2007ih, Endo:2007sz, Watanabe:2010vy}, through the conformal (trace) anomaly, $\phi$ acquires an effective interaction with the gauge boson sector, which depends on the running of the gauge coupling:
\begin{align}
\label{eq: effective interaction from trace anomaly}
\mathcal{L}_{\mathrm{eff}} \supset
\sqrt{-g}\, \frac{- \Lambda \phi}{M_{\mathrm{Pl}}^2} \Bigg(
   \frac{\beta(g')}{2 g'} B_{\mu\nu} B^{\mu\nu} 
 &+ \frac{\beta(g)}{2 g} W_{\mu\nu}^a W^{a\,\mu\nu} 
 \nonumber \\
 &+ \frac{\beta(g_s)}{2 g_s} G_{\mu\nu}^a G^{a\,\mu\nu} 
\Bigg),
\end{align}
where $\beta(g')$ denotes the beta function of $U(1)_Y$ hypercharge gauge coupling $g'$, while $\beta(g)$ and $\beta(g_s)$ denote the beta function for $SU(2)_L$ weak coupling $g$ and $SU(3)_c$ strong coupling $g_s$, respectively.
In appendix~\ref{app: action in d-dimension} and ~\ref{app: interactions of secluded scalar}, we show the derivation of Eqs.~\eqref{eq: mixing between phi and Higgs} and ~\eqref{eq: effective interaction from trace anomaly} from the action in $d$ dimensions.

In addition to these interactions, we assume that in the Einstein frame, the action of the dark glueball $\phi$ includes the following effective interaction described by a higher-dimensional operator:\footnote{
Even if this type of interaction is absent in the original action, it generally emerges at low energies as an effective interaction when particles heavier than the scalar field $\phi$ exist in the dark sector~\cite{Ema:2021fdz}.
}
\begin{align}
\label{eq: effective interaction with quadratic curvature}
\mc{L}_{\mathrm{eff}} \supset - c_{\phi RR} \frac{\phi}{\Lambda} R_{\mu\nu\rho\sigma} R^{\mu\nu\rho\sigma},
\end{align}
where $c_{\phi RR}$ is the coupling between $\phi$ and the quadratic curvature (the square of the Riemann tensor).
Since this coupling $c_{\phi RR}$ strongly depends on the hidden sector in the model, we treat $c_{\phi RR}$ as a free parameter in our analysis. As we will discuss later, this interaction induces the decay of $\phi$ into gravitons.\footnote{
Other types of couplings to the quadratic curvature, such as $\phi R^2$ and $\phi R^{\mu\nu} R_{\mu\nu}$, do not induce the decay into gravitons in the flat spacetime~\cite{Delbourgo:2000nq, Alonzo-Artiles:2021mym, Ema:2021fdz}.
}

\subsection{Palatini formulation}
In the Palatini formalism, after the Weyl transformation~\eqref{eq: weyl transformation}, the Ricci scalar transforms as 
\begin{align}
\hat{R} = \Omega^{2} R.
\end{align}

Similar to the case of the metric formalism, the Weyl transformation generates interactions between $\phi$ and the Higgs sector. However, in the Palatini formulation, the Ricci scalar transforms differently, and no additional derivative terms for $\phi$ are induced. 
As a consequence, the effective interactions between $\phi$ and the SM fields take a simpler form. 
In particular, after expanding $\Omega$ around the vacuum, we obtain
\begin{align}
S \supset 
    \int d^4 x  {- 2 \Lambda \over M_{\mr{pl}}^2} 
     \phi |\partial_{\mu} H|^2 .
\end{align}

Or equivalently, in terms of canonically normalized Higgs field by $H_c$, we can express the three-point interaction as
\begin{align}
\label{eq: mixing between phi and Higgs in Palatini}
S \supset 
    \int d^4 x {- \Lambda \over M_{\mr{pl}}^2} 
      (\Box \phi) |H_c|^2.
\end{align}
These interaction terms correspond to those in metric formalism,~\eqref{eq: mixing between phi and Higgs} and~\eqref{eq: mixing between phi and canonical Higgs}, with $\xi =0$.

For the couplings between the gauge bosons and the quadratic curvature terms (the square of the Riemann tensor), we can treat them in the same manner as in the metric formalism, by considering the effective interactions given by Eqs.~\eqref{eq: effective interaction from trace anomaly} and~\eqref{eq: effective interaction with quadratic curvature}. Therefore, hereafter, we simply present the results in the metric formalism. The case of the Palatini formalism can be understood as the limit with $\xi = 0$.

%%%%%%%%%%%%%%%%%%%%%%%%%%%%%%%%%%%%%%%%
\section{Secluded scalar decay by conformal symmetry breaking}
\label{sec: secluded scalar decay}
%%%%%%%%%%%%%%%%%%%%%%%%%%%%%%%%%%%%%%%%

In this section, we summarize the primary decay channels of the secluded scalar field $\phi$ into SM particles.  
At tree level, the main decay channel is the two-body decay into Higgs bosons through the interaction given in Eq.~\eqref{eq: mixing between phi and canonical Higgs}.  
This decay rate depends on $\xi$ and becomes dominant when $\xi$ is sufficiently large; conversely, it vanishes in the conformal limit $\xi = 1/6$.  
The leading contribution to the decay rate that does not depend on $\xi$ is the two-body decays into gauge bosons induced by the conformal (trace) anomaly given in Eq.~\eqref{eq: effective interaction from trace anomaly}. Among them, the decay into gluons is particularly dominant.

\subsection{Decay into Higgs bosons}
Through the interaction between $\phi$ and the Higgs field induced by the non-minimal coupling~\eqref{eq: mixing between phi and canonical Higgs}, $\phi$ decays into the Higgs bosons. This decay rate is given by
\begin{align}
\label{eq: Decay rate into Higgs}
\Gamma(\phi \to H_c H_c^{*})
    = {1 \over 8 \pi} (1- 6\xi)^2 \left(\Lambda \over M_{\mr{pl}}^2\right)^2 m_{\phi}^3.
\end{align}
For simplicity, we neglect phase space suppression here and in what follows.
At the tree level, there are no other available two-body decay channels for $\phi$, 
and thus the decay into the Higgs bosons is one of the primary channels to SM particles. 

\subsection{Decay into Gauge bosons}
The effective interaction given in Eq.~\eqref{eq: effective interaction from trace anomaly} 
allows for two-body decays of $\phi$ into gauge bosons~\cite{Endo:2007ih, Endo:2007sz, Watanabe:2010vy}. 
The anomaly-induced decay rates of $\phi$ into gauge bosons are given by
\begin{align}
\label{eq: Decay rate into BB}
\Gamma_{\mathrm{AI}}(\phi \to B_{\mu} B_{\nu}) 
    &= \frac{\alpha^{\prime 2}}{256 \pi^3}
       \left(\frac{\Lambda}{M_{\mathrm{pl}}^2}\right)^2
       m_{\phi}^3 
       \left|\sum_{i = s, f, v} b_i^{U(1)_Y}\right|^2, \\
\label{eq: Decay rate into WW}
\Gamma_{\mathrm{AI}}(\phi \to W^a_{\mu} W^a_{\nu}) 
    &= \frac{\alpha^{2}}{256 \pi^3}
       \left(\frac{\Lambda}{M_{\mathrm{pl}}^2}\right)^2
       m_{\phi}^3 
       \left|\sum_{i = s, f, v} b_i^{SU(2)}\right|^2, \\
\label{eq: Decay rate into gluons}
\Gamma_{\mathrm{AI}}(\phi \to G^a_{\mu} G^a_{\nu}) 
    &= \frac{\alpha_s^{2}}{256 \pi^3}
       \left(\frac{\Lambda}{M_{\mathrm{pl}}^2}\right)^2
       m_{\phi}^3 
       \left|\sum_{i = s, f, v} b_i^{SU(3)}\right|^2,
\end{align}
where $\alpha'$,$\alpha$, and $\alpha_s$ are the fine structure constants for each gauge coupling, $g'$, $g$, and $g_s$, respectively, and $\sum_i b_i^{\cdots}$ are the coefficients determined by the beta functions, as we will show below. We note that the index $a$ is not summed over in these equations.

For the case of $SU(N)$ gauge interactions, these coefficients and beta functions are related by
\begin{align}
&\beta(g) = - {g^3 \over (4 \pi)^2} \sum_{i = s, f, v}b_i^{SU(N)}  + \mc{O}(g^5),\\
&\sum_{i = s, v, f} b_i^{SU(N)} =  \left[
    {11 \over 3}N -{1 \over 6}N_s - {1 \over 3}N_f \right],
\end{align}
where $N_s$ and $N_f$ are internal quantum numbers of (real) scalar and (Weyl) fermion, which are charged under $SU(N)$ interaction.  For the case of $U(1)$ gauge interaction,
\begin{align}
&\beta(g') = - {g^3 \over (4 \pi)^2} \sum_{i = s, f, v}b_i^{U(1)}  + \mc{O}(g^5),\\
&\sum_{i = s, v, f} b_i^{U(1)} = \left[
    -{Q_s^2 \over 3}N'_s - {2 Q_f^2 \over 3}N'_f \right],
\end{align}
where $N'_s$ and $N'_f$ are internal quantum numbers of (real) scalar and (Weyl) fermion, which are charged under $U(N)$ interaction, and $Q_s$ and $Q_f$ are their charges, respectively.

In the case of SM, these are given by
\begin{align}
&\sum_{i = s, f, v}b_i^{SU(3)_c} = 7,\\
&\sum_{i = s, f, v}b_i^{SU(2)_L} = 19/6, \\
&\sum_{i = s, f, v}b_i^{U(1)_Y} = -41/6.
\end{align}

Importantly, these anomaly-induced decay rates are independent of the Higgs non-minimal coupling constant $\xi$. 
Therefore, near the conformal limit $\xi \simeq 1/6$, the decay into Higgs bosons~\eqref{eq: Decay rate into Higgs} is suppressed while anomaly-induced processes remain and become the leading decay channel. Among the decay channels into the gauge bosons, the decay into gluons has the largest decay rate due to the large coupling constant and the coefficient of the beta function.

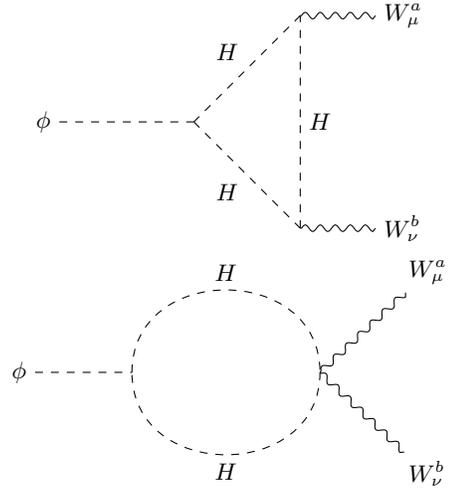
\begin{figure}[h]
\centering
\begin{tikzpicture}
  \begin{feynman}
    \vertex (phi) {\(\phi\)};
    \vertex [right=2cm of phi] (v0);
    
    \vertex [above right=2cm of v0] (v1);
    \vertex [below right=2cm of v0] (v2);
    
    \vertex [right =1cm of v1] (A1) {\(W^{a}_\mu\)};
    \vertex [right =1cm of v2] (A2) {\(W^{b}_\nu\)};

    \diagram* {
      (phi) -- [scalar] (v0),
      (v1) -- [photon] (A1),
      (v2) -- [photon] (A2),

      (v0) -- [scalar, edge label'=\(H\)] 
      (v2) -- [scalar, edge label'=\(H\)] 
      (v1) -- [scalar, edge label'=\(H\)] (v0),
    };
  \end{feynman}
  \end{tikzpicture}
  \begin{tikzpicture}
  \begin{feynman}
    \vertex (phi) {\(\phi\)};
    \vertex [right=1.5cm of phi] (a);
    \vertex [right=4cm of phi] (b);
    \vertex [above right =1.5cm of b] (A1) {\(W^{a}_\mu\)};
    \vertex [below right =1.5cm of b] (A2) {\(W^{b}_\nu\)};

    \diagram*{
      (phi) -- [scalar] (a),

      (a) -- [scalar, half left,  looseness=1.5, edge label=\(H\)] (b),
      (b) -- [scalar, half left, looseness=1.5, edge label=\(H\)] (a),
      (b) -- [photon] (A1),
      (b) -- [photon] (A2),
    };
  \end{feynman}
\end{tikzpicture}
\caption{Higgs loop induced diagram}
\label{fig: higgs loop induced diagram}
\end{figure}

Another contribution to decay into gauge bosons arises from the loop diagram with the three-point interaction between $\phi$ and the Higgs field~\eqref{eq: mixing between phi and canonical Higgs}. We show the corresponding diagrams in Fig.~\ref{fig: higgs loop induced diagram}. 
These Higgs loop–induced contributions yield the following decay rates into the weak gauge boson $W_{\mu}$ and the hypercharge gauge boson $B_{\mu}$:
\begin{align}
\Gamma_{\mathrm{HLI}}(\phi \to B_{\mu} B_{\nu}) 
	&= \frac{\alpha'^2}{64 \pi^3 }\,
	   \left(\frac{\Lambda}{M_{\rm Pl}^2}\right)^2
       (1- 6\xi)^2 m_{\phi}^3, \\
\Gamma_{\mathrm{HLI}}(\phi \to W^a_{\mu} W^a_{\nu}) 
	&=  \frac{\alpha^2}{1024 \pi^3} 
	    \left(\frac{\Lambda}{M_{\rm Pl}^2}\right)^2
       (1- 6\xi)^2 m_{\phi}^3.
\end{align} 
Their impact on the total decay rate into the SM particles except for the graviton is always negligible, because these contributions are subdominant compared to the decay channel into Higgs bosons~\eqref{eq: Decay rate into Higgs}.

%%%%%%%%%%%%%%%%%%%%%%%%%%%%%%%%%%%%%%%%%%%%%%%%%%%%%%%%%%%%%%
%%%%%%%%%%%%%%%%%%%%%%%%%%%%%%%%%%%%%%%%%%%%%%%%%%%%%%%%%%%%%%

\subsection{Comments on decay into fermions}

Finally, we make some comments on decay channels involving SM fermions. In our setup, $\phi$ does not have a direct interaction with massless fermions in the Einstein frame. (See Appendix~\ref{app: action in d-dimension} for details.) Therefore, the two-body decays of $\phi$ into the SM fermions are forbidden at the tree level before the electroweak phase transition.  

\begin{figure}[h]
\centering
\begin{tikzpicture}
  \begin{feynman}
    \vertex (a) {\(\phi\)};
    \vertex [right=2cm of a] (b);
    \vertex [above right=1.5cm of b] (f1) {\(H\)};
    \vertex [below right=1.2cm of b] (c);
    \vertex [above right=1.2cm of c] (f2) {\(\psi\)};
    \vertex [below right=1.2cm of c] (f3) {\(\psi\)};
    
    \diagram* {
      (a) -- [scalar] (b) -- [scalar] (f1),
      (b) -- [scalar, edge label'=\(H\)] (c),
      (c) -- [fermion] (f2),
      (c) -- [anti fermion] (f3),
    };
  \end{feynman}
\end{tikzpicture}
\caption{Decay channel involving SM fermions $\psi$ at the tree level}
\label{fig: decay channel involving fermions}
\end{figure}
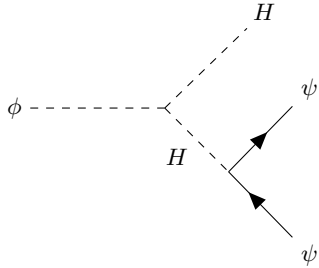

Decays into fermions may occur through three-body decays $\phi \to H_c \psi_{Lc} \psi_{Rc}$ as illustrated in Fig.~\ref{fig: decay channel involving fermions} at the tree level, where $\psi_{L(R)c}$ denotes the canonically normalized SM fermions. However, their decay widths are proportional to $(1 - 6\xi)^2$ because they include three-point interaction between $\phi$ and the Higgs field~\eqref{eq: mixing between phi and canonical Higgs} and they are always subdominant compared to the decay rate into Higgs bosons~\eqref{eq: Decay rate into Higgs}.

At higher order, however, we note that analogously to Eq.~\eqref{eq: effective interaction from trace anomaly}, we should have the contribution to the effective Lagrangian which depends on the running of the Yukawa coupling
\begin{align}
\label{eq: another effective interaction from trace anomaly}
\mathcal{L}_{\mathrm{eff}} \supset \sqrt{-g}\, \frac{\Lambda \phi}{M_{\mathrm{Pl}}^2} \beta(y_\psi) H_c \psi_{Lc} \psi_{Rc} + {\rm h.c.},
\end{align}
where $\beta(y_{\psi})$ denotes the beta function of Yukawa coupling $y_{\psi}$.

This contribution, as in Eq.~\eqref{eq: effective interaction from trace anomaly}, is independent of $\xi$ and induces the three-body decay of $\phi$. However, it is always subdominant compared to the two-body decay into gluons. This is because, even after taking into account the running effects, $\alpha_t < \alpha_s$ where $\alpha_t:= y_t^2 / (4\pi^2)$ with $y_t$ being the top Yukawa coupling, and the three-body decay is suppressed by the phase space factor $1/4 \pi$ relative to the two-body decay.

After the electroweak phase transition, since the fermion mass breaks conformal symmetry, a two-body decay channel into fermions becomes allowed. However, the rate is strongly suppressed by the ratio of the fermion mass to the scalar mass, 
$(m_{\psi} / m_{\phi})^2$. This mass suppression cannot be avoided by attaching an additional Yukawa vertex and considering a three-body decay.

%%%%%%%%%%%%%%%%%%%%%%%%%%%%%%%%%%%%%%%%%%%%%%%%%%%%%%%%%%%%%%
%%%%%%%%%%%%%%%%%%%%%%%%%%%%%%%%%%%%%%%%%%%%%%%%%%%%%%%%%%%%%%

%%%%%%%%%%%%%%%%%%%%%%%%%%%%%%%%%%%%%%%%
\section{Graviton as dark radiation}
\label{sec: graviton as dark radiation}
%%%%%%%%%%%%%%%%%%%%%%%%%%%%%%%%%%%%%%%%
In this section, we analyze and discuss the graviton production associated with the decay of $\phi$ and the constraints from dark radiation. 
The secluded scalar field, which interacts only through gravity, can generally decay into gravitons. 
Indeed, the interaction in Eq.~\eqref{eq: effective interaction with quadratic curvature} induces the dark glueball decay into gravitons with the following decay rate~\cite{Ema:2021fdz} :
\begin{align}
\label{eq: Decay rate into gravitons}
\Gamma(\phi \to 2g) = \frac{c_{\phi RR}^2}{4\pi} \frac{m_{\phi}^7}{ \Lambda^2 M_{\mathrm{pl}}^4}.
\end{align}

If this decay rate into gravitons is comparable to or larger than that into the SM particles, and if the scalar accounts for a dominant or at least non-negligible fraction of the total energy density at the time of decay, a large number of high-frequency gravitons are produced~\cite{Strumia:2025dfn}. These gravitons behave as dark radiation and can affect the thermal evolution of the universe, thereby influencing cosmological observables such as the CMB spectrum~\cite{Smith:2006nka, Sendra:2012wh, Pagano:2015hma, Clarke:2020bil}. In the following, for simplicity, we assume that the scalar $\phi$ dominates the universe at the time of decay.
This is the case, for example, if $\phi$ is an elementary field produced by the misalignment mechanism with an initially large amplitude, or if the secluded dark gauge sector is reheated to a temperature comparable to or higher than that of the SM sector.

To qualitatively understand this impact, it is important to consider the ratio of the decay rate into gravitons to the total decay width $\Gamma_{\phi}$, defined as
\begin{align}
B_{g} := \frac{\Gamma(\phi \to 2g)}{\Gamma_{\phi}},
\end{align}
as well as the temperature at which $\phi$ decays
\begin{align}
T_{\mr{dec}} \simeq \left(90 (1-B_g) \over \pi^2 g_{*}(T_{\mr{dec}})\right)^{1/4} \sqrt{\Gamma_{\phi}M_{\rm pl}},
\end{align}
where  $g_{*}(T_{\mathrm{dec}})$ represents the effective number of relativistic degrees of freedom at $T_{\mr{dec}}$. Here we assume that the decay into gravitons is a subdominant decay channel, $B_g < 1$.

As discussed in the previous section, the dominant decay channels into the SM particles are the two-body decays into the Higgs bosons and gluons. Therefore, $B_g$ can be approximately expressed as
\begin{align}
\label{eq: branching ratio}
B_g \simeq \frac{r^2}{r^2 + \Delta \xi^2 + \frac{49 
\, \alpha_s^2}{4\pi^2}},
\end{align}
where we define $\Delta \xi := |1-6 \xi|$, and the strong gauge coupling is evaluated at the scale of the scalar mass.
Here we introduce a dimensionless parameter $r$ to normalize the decay rate into gravitons,
\begin{align}
\label{eq: r definition}
r := \sqrt{2} c_{\phi RR}\, \frac{m_{\phi}^2}{\Lambda^2} .
\end{align}
In the dark glueball case, we expect $r = {\cal O}(1)$ since $m_\phi \sim \Lambda$ and $c_{\phi RR} = {\cal O}(1)$.\footnote{\label{note: relation between mass and lambda}
In pure SU(3) Yang--Mills theory, lattice simulations indicate that the lightest
scalar glueball mass is $m_{0^{++}} \simeq (6\text{--}7)\,\Lambda_{\overline{\rm MS}}$~\cite{Ishikawa:2017xam,Athenodorou:2020ani}.
}
This leads to $B_g = {\cal O}(0.1)$ as long as $|\xi| = {\cal O}(1)$. 
In particular, for $\xi \simeq \frac{1}{6}$, we have $B_g \simeq 1$.
Thus, contrary to the naive expectation in Ref.~\cite{Strumia:2025dfn}, the production of gravitons from the dark glueball decay is generically sizable. This difference arises because the dominant decay channel into SM particles is the decay into Higgs bosons, so we do not benefit from the enhancement associated with the large number of SM degrees of freedom. In particular, the decays into gauge bosons and fermions are suppressed either by a loop factor or by the small fermion masses.

For a given $B_g$, the contribution from dark radiation to the effective number of neutrinos is expressed as
\begin{align}
\Delta N_{\mathrm{eff}} 
= \frac{43}{7} 
\left(\frac{43/4}{g_{*}(T_{\mathrm{dec}})}\right)^{1/3} 
\frac{B_g}{1 - B_g}.
\end{align}
The current CMB+BAO observations provide upper bounds on this observable quantity as
\begin{align}
\Delta N_{\mathrm{eff}}  
	\leq  
\begin{cases}
0.11 & (\text{Planck 2018~\cite{Planck:2018vyg}}) ,\\
0.19 & (\text{DESI 2024~\cite{DESI:2024hhd}}),
\end{cases}
\end{align}
which provide upper limits on $B_g$, and consequently, constraints on the parameter space $(r, \Delta \xi)$.

\begin{figure}[t]
  \centering
  \includegraphics[width=0.38\textwidth]{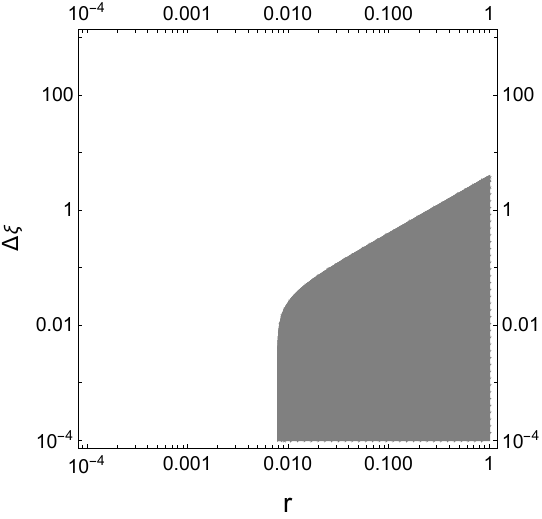}
  \caption{Constraints on the graviton dark radiation in the $(r, \Delta \xi)$ plane, where $\Delta \xi := |1 - 6\xi|$ and $r$ is defined in Eq.~\eqref{eq: r definition}. The shaded region is excluded because it violates the bound $\Delta N_{\mathrm{eff}} \lesssim 0.19$ derived from DESI observations~\cite{DESI:2024hhd}. We have adopted $m_{\phi} = 3 \times 10^{13}~\mr{GeV}$  and $\Lambda = m_{\phi}/6$ for the benchmark values for which $T_{\mathrm{dec}} \simeq 420~\mathrm{GeV}$ even in the case of $(r, \Delta \xi) = (0, 0)$. 
  For larger $m_\phi$, the left vertical boundary  shifts slightly to smaller $r$.
}
  \label{fig: Excluded region DR}
\end{figure}

We show in Fig.~\ref{fig: Excluded region DR} the constraint $\Delta N_{\mathrm{eff}}  \lesssim 0.19$~\cite{DESI:2024hhd} represented on the $(r, \Delta \xi)$ plane for $m_{\phi} = 3 \times 10^{13}~\mathrm{GeV}$,where the shaded region is excluded. Here we take $\Lambda = m_{\phi}/6$ based on the lattice results of pure SU(3) Yang--Mills theory. (See footnote~\ref{note: relation between mass and lambda}.)
Under these choices, even for $(r, \Delta \xi) = (0, 0)$, the decay temperature is as high as $T_{\mathrm{dec}} \simeq 420~\mathrm{GeV}$, and consequently we can take $g_{*}(T_{\mathrm{dec}}) = 106.75$ uniformly over the allowed region in $(r, \Delta \xi)$ plane.
%which justifies the analysis assuming that the Higgs field has no vacuum expectation value. 
We adopt the reference value of $m_\phi$ to evaluate $\alpha_s$, which determines the limit on $r$ for $\Delta \xi \ll 1$; for larger $m_\phi$, the left vertical boundary of the excluded region shifts slightly to smaller $r$.

The region $0.01 \lesssim \Delta \xi < r$ is excluded because it violates the upper bound on $\Delta N_{\mathrm{eff}}$, while if the Higgs field has a sufficiently large non-minimal coupling to gravity, this constraint can be avoided because the main decay channel is into the SM Higgs bosons and the branching fraction of the gravitons is suppressed. On the other hand, when both $r$ and $\Delta \xi$ are small, the region remains allowed even if $\Delta \xi < r$, because in this region, the decays into gauge bosons become dominant.
In the dark glueball case, we expect $r = {\cal O}(1)$, as discussed above. Consequently, the dark-radiation bound requires a sizable non-minimal coupling of the SM Higgs, $\Delta \xi \gtrsim {\cal O}(1)$.

%%%%%%%%%%%%%%%%%%%%%%%%%%%%%%%%%%%%%%%%%%%%%%%%%%%%%%%%%%%%%%%%%%%%
%%%%%%%%%%%%%%%%%%%%%%%%%%%%%%%%%%%%%%%%%%%%%%%%%%%%%%%%%%%%%%%%%%%%%

%%%%%%%%%%%%%%%%%%%%%%%%%%%%%%%%%%%%%%%%
\section{Stochastic gravitational wave spectrum }
\label{sec: stochastic gravitational wave spectrum}
%%%%%%%%%%%%%%%%%%%%%%%%%%%%%%%%%%%%%%%%

Gravitons produced by the decay of $\phi$ in the early universe exhibit a characteristic spectrum with a peak in the high-frequency range.  The stochastic gravitational wave spectrum from these gravitons in the present day  is given by~\cite{Ema:2021fdz}
\begin{align}
\label{eq: GW spectrum}
{d\Omega_{\rm GW} \over d \ln E} ={1 \over \rho_{\mr{crit}}} {16 E^4 \over m_{\phi}^4} 
{B_g \Gamma_{\phi} \, \rho_{\phi} (z_d) \over H(z_d)},
\end{align}
where $1 + z_d := m_{\phi}/2E$ and $\rho_{\mr{crit}}$ denotes the critical energy density. $H(z)$ and $\rho_{\phi}(z)$ are the Hubble parameter and the energy density of $\phi$ at redshift $z$, respectively.

\begin{figure}[t]
  \centering
  \includegraphics[width=0.45\textwidth]{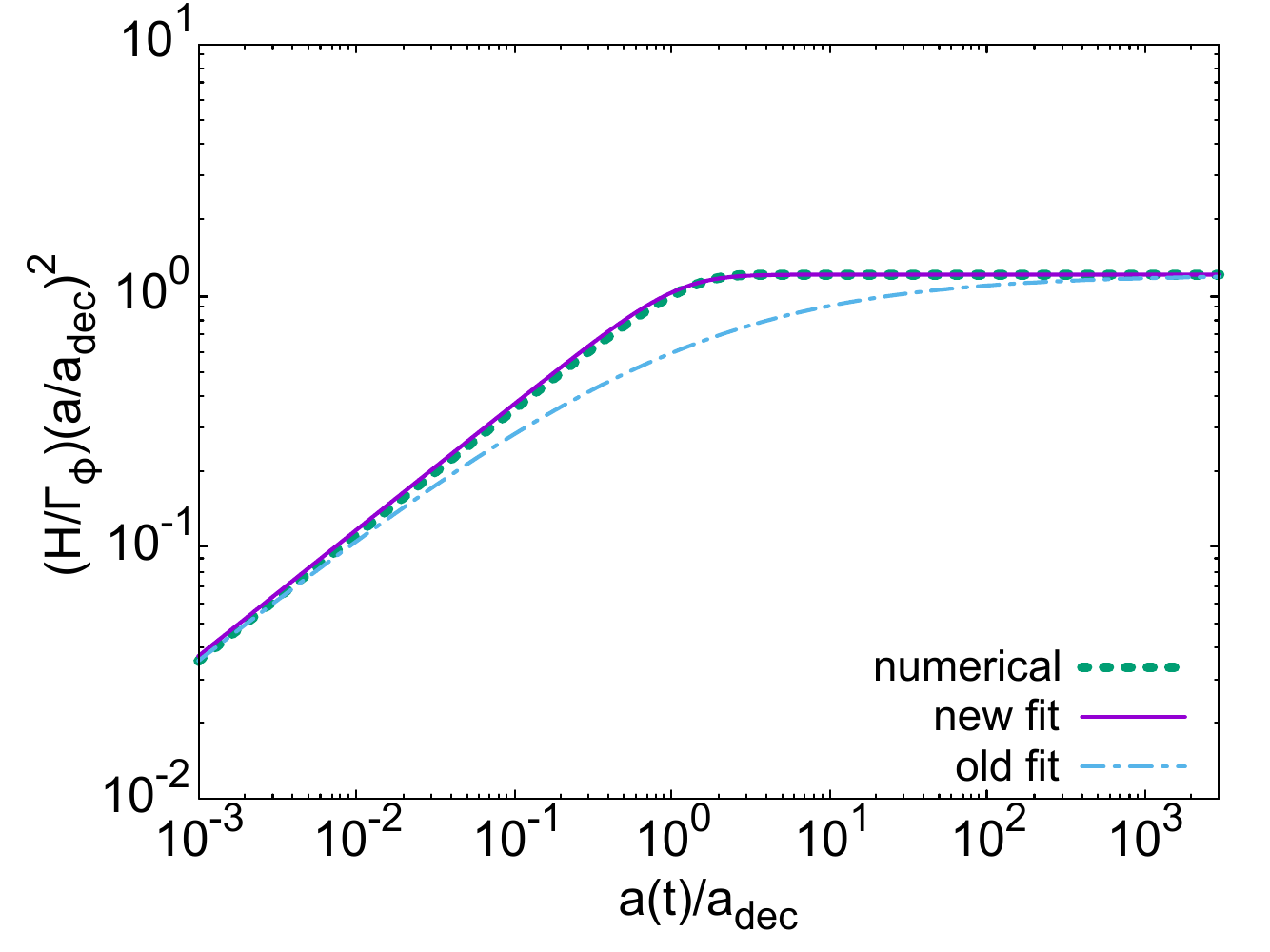}
  \caption{Comparison of $H(z)$ between the numerical result (green short dashed line, ``numerical'') and the new fitting formula (\ref{eq: fitting function of Hubble}) (solid line, ``new fit'') as well as the old fitting formula provided in Ref.~\cite{Mudrunka:2023wxy} (blue long dashed line, ``old fit'').
  Here we defined $a_{\rm dec}\equiv a(H=\Gamma_\phi)$.}
  \label{fig:H_fit}
\end{figure}

In order to evaluate the GW spectrum (\ref{eq: GW spectrum}), we need to know the Hubble parameter $H(z)$, the cosmic time $t(z)$ and the $\phi$ energy density $\rho_\phi(z)$ as a function of redshift $z$. 
In the case of our interest, $\phi$ dominates the universe before it decays and hence it will be convenient to have a fitting formula for these quantities that connects the $\phi$- (or matter-)dominated and radiation-dominated era. 
Here we provide useful fitting formulae:
\begin{align}
	&t(z) = t_{\rm ref}\left(\frac{a}{a_{\rm ref}}\right)^2 \left[ 1+ c n\left(\frac{H_{\rm ref}}{\Gamma_{\phi}}\right)^{\frac{n}{2}}
	\left(\frac{a_{\rm ref}}{a}\right)^n  \right]^{\frac{1}{2n}}, 
    \label{eq: approximate expression of time}\\
    &H(z) = \frac{1}{a \frac{dt}{da}} = \frac{1}{t(z)} \frac{2+2cn \left(\frac{H_{\rm ref}}{\Gamma_{\phi}}\right)^{\frac{n}{2}} \left(\frac{a_{\rm ref}}{a}\right)^n}{4+3cn \left(\frac{H_{\rm ref}}{\Gamma_{\phi}}\right)^{\frac{n}{2}} \left(\frac{a_{\rm ref}}{a}\right)^n},
    \label{eq: fitting function of Hubble}
\end{align}
where $c$ and $n$ are parameters to be determined to fit the numerical results and the subscript ``ref'' means that the corresponding quantity is evaluated at the arbitrary reference time $t_{\rm ref}$ well after the $\phi$ decays, at which the universe is dominated by radiation; $t_{\rm ref}\gg \Gamma_\phi^{-1}$ or $T_{\rm{ref}} \ll T_{\rm{dec}}$,and we assume that the effective number of relativistic degrees of freedom does not change significantly between $T_{\rm{dec}}$ and $T_{\rm{ref}} $.\footnote{
Temperature and scale factor at the reference point are related to those of the present universe, $T_0$ and $a_0$, through 
\begin{align}
g_{*S}(T_{\mr{ref}}) T_{\mr{ref}}^3 a_{\mr{ref}}^3 = g_{*S}(T_0) T_{0}^3 a_0^3,
\end{align}
where $g_{*S}$ is the entropy effective degrees of freedom.
}
Therefore, $H_{\mr{ref}}$ and $t_{\rm ref}$ can be expressed as
\begin{align}
H_{\mr{ref}} =\sqrt{\pi^2 g_{*}(T_{\mr{ref}}) T_{\mr{ref}}^4 \over 90 M^2_{\mr{pl}}},~~~t_{\rm ref} = \frac{1}{2H_{\rm ref}}.
\end{align} 
We find that numerical results are well reproduced for $n=2$ and $c=1.5$.
Fig.~\ref{fig:H_fit} shows the comparison of $H(z)$ between the numerical result and the new fitting formula (\ref{eq: fitting function of Hubble}) (``new fit''), as well as the old fitting formula provided in Ref.~\cite{Mudrunka:2023wxy} (``old fit''). It is seen that the new fitting formula coincides with the numerical result very well and it is hard to see the difference.
Therefore, this fitting function is quite useful for obtaining the gravitational spectrum analytically.

%%%%%%%%%%%%%%%%%%%%%%%%%%%%%%%%%%%%%%%%%%%%%%%%%%%%%%%%%%%%%%%%%%%%
%%%%%%%%%%%%%%%%%%%%%%%%%%%%%%%%%%%%%%%%%%%%%%%%%%%%%%%%%%%%%%%%%%%%%

In the approximation where the energy density of the SM sector is neglected, $\rho_{\phi}$ can be expressed as
\footnote{
By explicitly including the dependence on the radiation energy density $\rho_{\mathrm{rad}}(z)$, $\rho_{\phi}$ can be expressed as
\begin{align}
\rho_{\phi}(z)
	\simeq \left( 1 + R(\tilde{t}_{\mathrm{ref}}) \right)^{-1}
	3 \tilde H_{\mathrm{ref}}^2 M_{\mathrm{pl}}^2 
	\left( \frac{\tilde a_{\mathrm{ref}}}{a(z)} \right)^3
	e^{- \Gamma_{\phi} t(z)} ,
\end{align}
where 
\begin{align}
R(\tilde{t}_{\mathrm{ref}}) := 
\frac{\rho_{\mathrm{rad}}(z(\tilde{t}_{\mathrm{ref}}))}{\rho_{\phi}(z(\tilde{t}_{\mathrm{ref}}))} <1,
\end{align}
represents the ratio of the radiation energy density to that of the $\phi$ field at $\tilde{t}_{\mathrm{ref}}$.}
\begin{align}
	\rho_\phi(z) 
    %&= \rho_\phi(\tilde t_{\rm ref})  \left(\frac{\tilde a_{\rm ref}}{a(z)}\right)^3 e^{-\Gamma_{\rm tot} t(z)} \nonumber \\
	&= 3 \tilde H_{\rm ref}^2 M_{\rm Pl}^2  \left(\frac{\tilde a_{\rm ref}}{a(z)}\right)^3 e^{-\Gamma_{\phi} t(z)} \nonumber \\
     &= \frac{16 H_{\rm ref}^2 M_{\rm Pl}^2}{3(cn)^{1/n}}\left(\frac{a_{\rm ref}}{a(z)}\right)^3 
	 \left(\frac{\Gamma_{\phi}}{H_{\rm ref}}\right)^{1/2} e^{-\Gamma_{\phi} t(z)},
     \label{eq:rhophiz}
\end{align}
where $t(z)$ is given by Eq.~\eqref{eq: approximate expression of time}, $\tilde H_{\rm ref}$ and $\tilde a_{\rm ref}$ are the Hubble parameter and scale factor at another reference time $\tilde{t}_{\mathrm{ref}} \ll \Gamma_{\phi}^{-1}$, i.e., much before the $\phi$ decay, so that the universe was $\phi$-dominated.
In the second line we used Eqs.~(\ref{eq: approximate expression of time}) and (\ref{eq: fitting function of Hubble}).

\begin{figure}[t]
  \centering
  \includegraphics[width=0.5\textwidth]{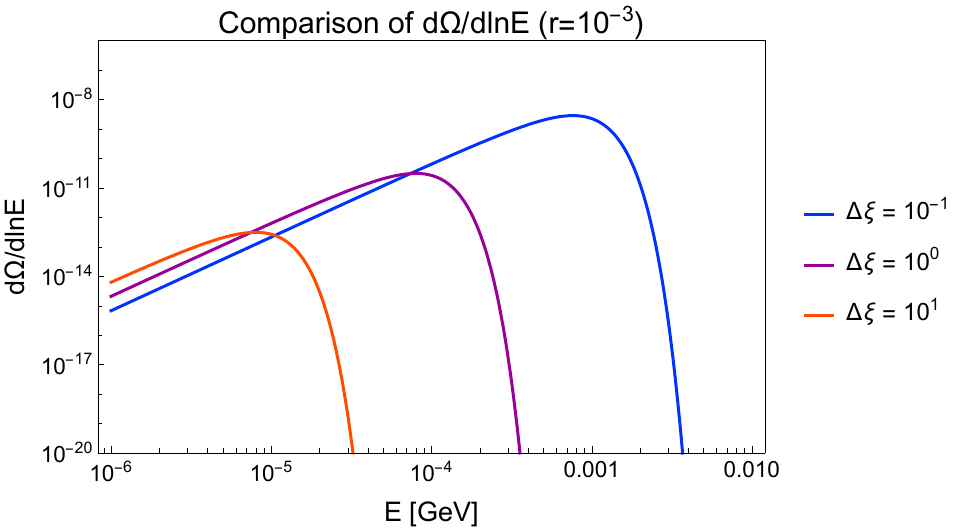}
   \par\vspace{1em}
  \includegraphics[width=0.5\textwidth]{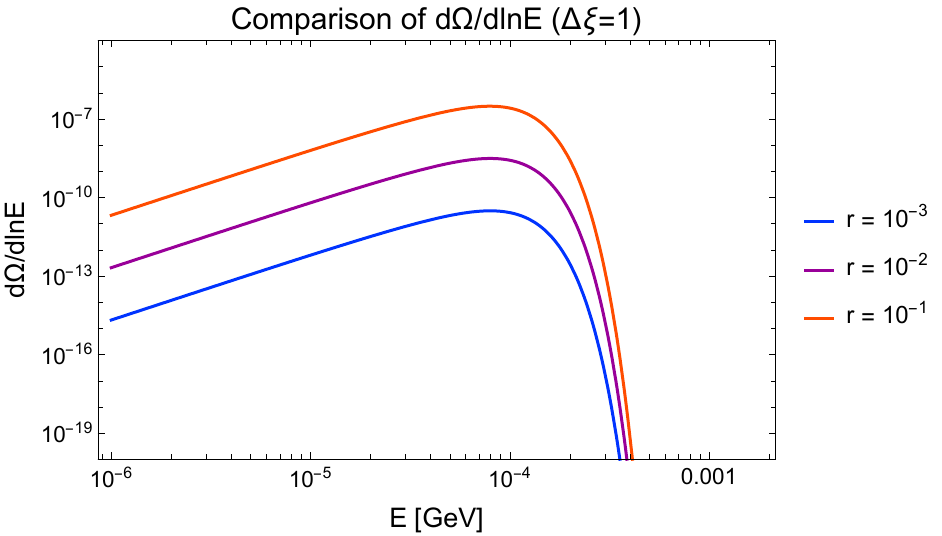}
   \par\vspace{1em}
  \includegraphics[width=0.5\textwidth]{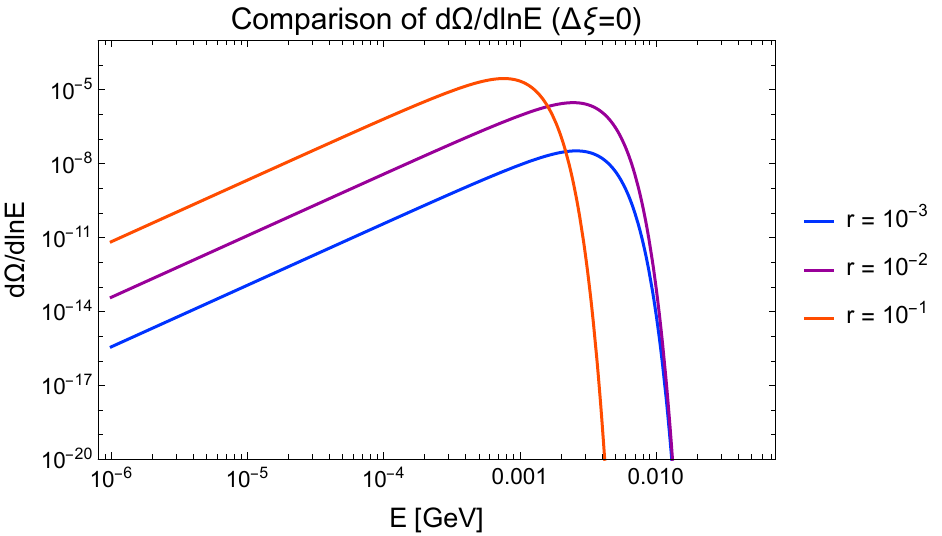}
  \caption{Comparison of the gravitational wave spectra for different combinations of $(r, \Delta \xi)$. 
  Top: spectra for varying $\Delta \xi$ with fixed $r = 10^{-3}$. Middle: spectra for varying $r$ with fixed $\Delta \xi = 1$. Bottom: spectra for varying $r$ with fixed $\Delta \xi = 0$. We have adopted $m_{\phi} = 3 \times 10^{13}~\mr{GeV}$ and $\Lambda = m_{\phi}/6$, as Fig.~\ref{fig: Excluded region DR}.
}
  \label{fig: GW spectrum}
\end{figure}

By substituting Eqs.~\eqref{eq:rhophiz} and \eqref{eq: fitting function of Hubble} into Eq.~\eqref{eq: GW spectrum}, the gravitational wave spectrum $d\Omega_{\mathrm{GW}} / d \ln E$ can be analytically computed in terms of the three input parameters $r$, $\Delta \xi$, and $m_{\phi}$. Note that all the reference time dependence disappears in the final results as it should be.\footnote{
     It is easily seen by noting that all the formulae are dependent on the combination $H_{\rm ref} a_{\rm ref}^2$, which is constant in time except for the change in the relativistic degrees of freedom.
}
We show in Fig.~\ref{fig: GW spectrum} the gravitational wave spectra compared among different combinations of $(r, \Delta \xi)$. We have adopted $m_{\phi} = 3 \times 10^{13}~\mr{GeV}$ and $\Lambda = m_{\phi}/6$ as Fig.~\ref{fig: Excluded region DR}.

Before proceeding, let us first comment on the general features that are common to all spectra. As is evident from Eq.~\eqref{eq: GW spectrum}, the gravitational wave spectrum depends on the $\rho_{\phi}$ at the time when gravitons were produced, as well as on the Hubble parameter at that epoch. In particular, if $\rho_{\phi}$ dominated the universe, the gravitational wave spectrum behaves as ${d\Omega_{\rm GW} / d \ln E} \propto E^{5/2}$ in the low-energy regime, and the peak energy is given by $E_{\mathrm{peak}} \simeq ({m_{\phi} / 2}) {T_{\mathrm{0}}/T_{\mathrm{dec}}}$~\cite{Ema:2021fdz, Mudrunka:2023wxy}.

%%%%%%%%%%%%%%%%%%%%%%%%%%%%%%%%%%%%%%%%%%%%%%%%%%%%%%%%%%%%%%%%%%%%
%%%%%%%%%%%%%%%%%%%%%%%%%%%%%%%%%%%%%%%%%%%%%%%%%%%%%%%%%%%%%%%%%%%%%

The top panel of Fig.~\ref{fig: GW spectrum} shows the spectra obtained by varying $\Delta \xi$ while fixing $r = 10^{-3}$. This corresponds to fixing the partial decay width to the gravitons $\Gamma (\phi \to gg)$ and varying the total decay rate $\Gamma_{\phi}$. Consequently, the low-energy tail of the spectrum remains almost unchanged, whereas only the position of the peak shifts. The middle panel corresponds to the spectra obtained by varying $r$ while fixing $\Delta \xi = 1$. In this case, since the decay into the SM particles dominates for the range of $r$ considered, this situation corresponds to fixing the total decay rate $\Gamma_{\phi}$ and varying the partial width $\Gamma (\phi \to gg)$. Consequently, the position of the peak remains unchanged, whereas only the amplitude of the spectrum increases or decreases. Finally, the bottom panel shows the spectra obtained by varying $r$ while fixing $\Delta \xi = 0$. In this case, both $\Gamma (\phi \to gg)$ and $\Gamma_{\phi}$ vary simultaneously, and therefore the resulting change in the spectra can be interpreted as a combination of the variations observed in the top and middle panels. We note that the parameter choice $(r, \Delta \xi) = (0.1, 0)$ does not satisfy the constraint from $\Delta N_{\mathrm{eff}}$, but it is shown in the figures as a benchmark to illustrate the behavior.

%%%%%%%%%%%%%%%%%%%%%%%
\section{Discussion and Conclusions}
\label{sec: discussion and conclusions}
%%%%%%%%%%%%%%%%%%%%%%%
Thus far, our analysis has focused on the ground state of the dark glueball, and we now discuss the cosmological impact of the other excitation states. We describe each state of the dark glueball by its global quantum numbers $J^{PC}$, where $P$ and $C$ denote the parity and charge-conjugation eigenvalues, respectively, and refer to the ground state as $0^{++}$.\footnote{See Ref.~\cite{Chen:2005mg, Meyer:2008tr, Athenodorou:2020ani} for lattice results of the glueball mass spectrum in $SU(N)$ gauge theories. For the classification of dark glueball states including gauge groups other than $SU(N)$, see Ref.~\cite{Gross:2020zam}.} 

We first consider the light C-even excited states. Assuming that gravity and the interactions in the dark sector does not violate CP, the interaction of the lightest P-odd excitation $\tilde{\phi} := 0^{-+}$ is restricted to the form
\begin{align}
\mathcal{L}_{\mathrm{eff}} \supset 
	-\, c_{\tilde{\phi} \tilde{R}R}\,
		{\tilde{\phi} \over \Lambda}\,
		\tilde{R}_{\mu\nu\rho\sigma} R^{\mu\nu\rho\sigma},
\end{align}
and hence it can decay only into gravitons. Therefore, its cosmological abundance may lead to slightly stronger constraints.
However, it should be noted that C-even states can eventually annihilate into the lightest $0^{++}$ mode via parametrically faster $2 \to 2$ processes~\cite{Forestell:2016qhc, Forestell:2017wov}.

Next, we consider the light C-odd states. These states are typically long-lived and, in particular, become stable if gravity and dark sector preserves the CP quantum number. Consequently, for a high confinement scale of the dark glueball, a stable C-odd state may easily lead to an overabundance. (See Refs.~\cite{Biondini:2024cpf,McKeen:2024trt} for the qualitative analysis of the relic abundance of the $1^{-+}$ state after the $0^{++}$ domination.) In such cases, one must either relax the assumption that gravity preserves CP or consider an alternative gauge group in which light C-odd states do not exist.

In our analysis, we have not assumed any Beyond Standard Model (BSM) field other than the secluded scalar. However, if there exist additional BSM fields that explicitly break conformal symmetry, their decay branching fractions should also be taken into account. As a representative example, let us consider right-handed neutrinos $N_i~(i=1,2,3)$ with masses $M_i$. The coupling between $N_i$ and $\phi$ is induced via the Majorana mass term of the right-handed neutrinos, and the corresponding decay rate is given by  
\begin{align}
\Gamma(\phi \to N_i N_i) \simeq \frac{1}{32\pi} \left(\frac{\Lambda}{M_{\mr{pl}}^2}\right)^2 \left(\frac{M_i}{m_{\phi}}\right)^2 m_{\phi}^3 \, .
\end{align}
This decay channel not only suppresses the production of graviton dark radiation but may also naturally lead to a successful non-thermal leptogenesis scenario~\cite{Endo:2006nj}, which we leave for future work.

In this work, we have investigated the parameter region consistent with observational constraints on graviton-induced dark radiation after the secluded scalar domination era. If secluded scalar $\phi$ couples only through gravity and dominates the early universe, the decay of $\phi$ can produce a large amount of high-frequency gravitons, and it can alter the cosmological observation, such as effective neutrino degrees of freedom. Paying attention to how the breaking of conformal invariance affects the amount of graviton dark radiation, we derived the corresponding viable parameter space. We have pointed out that when $\phi$ can decay into SM particles through a non-minimal coupling, its dominant decay channels are typically two-body decays into either Higgs bosons or gluons. In particular, the decay into Higgs bosons depends on the gravitational formulation (metric or Palatini). In the metric formalism, the branching ratio varies with the size of the Higgs non-minimal coupling $\xi$, and a sufficiently large $\xi$ can naturally alleviate the amount of graviton dark radiation. In contrast, in the Palatini formalism, the decay width into Higgs bosons is determined independently of the magnitude of $\xi$.

We then examined how the resulting gravitational-wave spectrum varies with the branching ratio into gravitons and the Higgs non-minimal coupling $\xi$. Our analytical calculations show that increasing the branching ratio into gravitons shifts the peak position to lower frequencies while enhancing its amplitude, whereas increasing $\xi$ likewise moves the peak to lower frequencies but reduces its height.

We note that although we have considered the dark glueball as a representative example of the secluded scalar, the same analysis applies equally to any scalar field that couples only through gravity, such as the (secluded) inflaton. In addition, although we have so far focused on gravitational decays of the scalar into SM particles, our results can of course also be applied when computing the decay rate into any additional sector beyond the Standard Model, if such a sector exists.

Before closing, we comment on some interesting implications of our results. We have seen that the decay of the secluded scalar field into SM Higgs bosons is the dominant channel if the Higgs non-minimal coupling $\xi$ is of order unity or larger. This implies that, even in the presence of additional hidden sectors, any secluded scalar would mainly decay into the SM sector if the non-minimal coupling of the Higgs is sufficiently large. In fact, for sufficiently large $\xi$, the same conclusion holds even if the hidden sector is not completely secluded and has only Planck-suppressed direct couplings to the SM or to other sectors, since the decay through the Higgs non-minimal coupling still dominates. A large non-minimal coupling of the SM Higgs also plays an important role in Higgs inflation~\cite{Bezrukov:2007ep}, where the Higgs inflaton naturally reheats the SM sector. In addition, such a large non-minimal coupling implies that, even if the universe is dominated by one or several dark sectors after inflation, the SM sector can be preferentially reheated by a dark scalar. In this case, the energy transferred to other hidden sectors is suppressed, which helps avoid an excessive abundance of dark radiation or dark matter originating from those sectors. Thus, a large non-minimal coupling of the SM Higgs may account for both cosmic inflation and the fact that mainly SM particles are populated in late-time cosmology.

%%%%%%%%%%%%%%%%%% Acknowledgements %%%%%%%%%%%%%%%%%%%%%%%%%%%%%%%%%%%%%%%
\section*{Acknowledgments}
%%%%%%%%%%%%%%%%%%%%%%%%%%%%%%%%%%%%%%%%%%%%%%%%%%%%%%%%%%%%%%%%%%%%%%%%%%%

This work was supported by JSPS KAKENHI (Grant Numbers 24K07010 [KN], 25H02165[FT], 25KJ0022 [JW]).
This work was also supported by World Premier International Research Center Initiative (WPI), MEXT, Japan, and is based upon work from COST Action COSMIC WISPers CA21106, supported by COST (European Cooperation in Science and Technology).

%%%%%%%%%%%%%%%%%%%%%%%%%%%%%%%%%%%%%%%%%%%%%%%%%%%%%%%%%%%
\appendix

\section{Action in d-dimension}
\label{app: action in d-dimension}
For the purpose of dimensional regularization, it is useful to write the action in $d$ dimensions in advance. In $d$ dimensions, the action can be written as
\begin{align}
S 
    &= \int d^d x \sqrt{-\hat{g}} \left[{M_{\mr{pl}}^{d-2} \over 2}\Omega^{d-2} \hat{R} + \mc{L}_{\phi} + \mc{L}_{SM} \right],\\
\Omega^{d-2} 
    & := 1 - {2 \Lambda^{(d-2)/ 2} \phi \over M_{\mr{pl}}^{d-2}} - {2 \xi |H|^2 \over M_{\mr{pl}}^{d-2}},
\end{align}
where the mass dimensions of $\phi$ and $H$ are $(d-2)/2$.

In the metric formalism, after the Weyl transformation $g_{\mu\nu} = \Omega^2 \hat{g}_{\mu\nu}$, the Ricci scalar transforms as~\cite{Wald:1984rg}
\begin{align}
\hat{R} 
    = &\Omega^{2} ( R + 2 (d-1) \Box \ln \Omega \nonumber \\
    &- (d-1)(d-2) g^{\mu\nu} \partial_{\mu} \ln \Omega \, \partial_{\nu} \ln \Omega ).
\end{align}
In the Palatini formalism, the transformation of the Ricci scalar is much simpler:
\begin{align}
\hat{R} = \Omega^{2} R.
\end{align}

Therefore, the action becomes
\begin{align}
\label{appeq: action in metric formalism}
S 
    = \int d^d x 
        &\sqrt{-g} \Bigg[{M_{\mr{pl}}^{d-2} \over 2} R + {1 \over \Omega^{d}} \left(\mc{L}_{\phi} + \mc{L}_{SM}\right) \nonumber \\
        &-  
            {(d-1)(d-2) \over 2} {1 \over \Omega^4}
            \frac{(\Lambda \partial_\mu\phi + \xi \partial_\mu |H|^2)^2}{M_{\rm Pl}^2}
    \Bigg],
\end{align}
in the metric formalism and 
\begin{align}
\label{appeq: action in Palatini formalism}
S 
    = \int d^d x 
        &\sqrt{-g} \Bigg[{M_{\mr{pl}}^{d-2} \over 2} R + {1 \over \Omega^{d}} \left(\mc{L}_{\phi} + \mc{L}_{SM}\right)
    \Bigg],
\end{align}
in the Palatini formalism.

\subsection{Higgs}
For a Higgs field, the action after the Weyl transformation takes the following form:
\begin{align}
\label{appeq: noncanonical Higgs action}
S 
    \supset \int d^d x \sqrt{-g} {1 \over \Omega^d}
        \Big[-\Omega^2 g^{\mu\nu}(D_{\mu} H^{\dagger}) (D_{\nu} H) - V(H) \Big].
\end{align}
Thus, the canonical scalar is defined as
\begin{align}
H_c := \Omega^{(2-d)/ 2} H.
\end{align}
In terms of the canonical field, the action becomes
\begin{align}
S 
    &\supset \int d^d x \sqrt{-g} 
        \Big[-g^{\mu\nu}(D_{\mu} H_c^{\dagger}) (D_{\nu} H_c)  -{1 \over \Omega^{d}} V(H) \nonumber \\
    &\qquad
       + \left({d-2 \over 2}\right)  
       \left\{{\Box\Omega\over\Omega} - {d \over 2} \left(\partial \Omega \over \Omega\right)^2\right\}|H_c|^2 \Big], \nonumber \\
    &= \int d^d x \sqrt{-g}
        \Bigg[-g^{\mu\nu}(D_{\mu} H_c^{\dagger}) (D_{\nu} H_c)  - {1 \over \Omega^{d}} V(H) \nonumber \\
\label{appeq: canonical Higgs action}
    &+
        \left({d-2 \over 2}\right)
        \left\{\left(4-d \over 4\right){\Box\Omega\over\Omega} - {d \over 4} 
        \left(\Box \Omega^{-1} \over \Omega^{-1}\right) \right\}|H_c|^2 \Bigg].
\end{align}

\subsection{Fermion}

For the fermion field, by noting that the vierbein is given by
\begin{align}
e_{\mu}^a = \Omega \hat{e}_{\mu}^a,
\end{align}
the action becomes 
\begin{align}
\label{appeq: noncanonical fermion action}
S \supset \int d^d x \sqrt{-g} {1 \over \Omega^{d-1}}
	\Bigg[- i \bar{\psi} e_a^{\mu} \gamma^a \hat{D}_{\mu} \psi \Bigg],
\end{align}
in the Einstein frame. Here, the covariant derivative includes the spin connection and the SM gauge interactions:
\begin{align}
 \hat{D}_{\mu} \psi : = (\partial_{\mu} -\hat{\Gamma}_{\mu} - ig A^a_{\mu} T^a).
\end{align}
The spin connection is given by
\begin{align}
	&\Gamma_\mu = -\frac{1}{4}\gamma_{[a} \gamma_{b]} \omega_\mu^{ab},\\
	&\omega_\mu^{ab} = \frac{1}{2}\big[
		e^{\nu a}(\partial_\mu e_\nu^b-\partial_\nu e_\mu^b)\nonumber \\
    &\qquad
        -e^{\nu b}(\partial_\mu e_\nu^a-\partial_\nu e_\mu^a)
		-(e^{\nu a}e^{\sigma b} - e^{\nu b}e^{\sigma a})e_{\mu}^c \partial_\nu e_{\sigma c}
	\big].
\end{align}
We note that the conformal transformation of the spin connection is
\begin{align}
	\hat\Gamma_\mu 
        &= \Gamma_\mu + \frac{1}{4}(\gamma^\nu\gamma_\mu-\gamma_\mu\gamma^\nu)\frac{\partial_\nu\Omega}{\Omega},
    \nonumber \\
	\gamma^\mu \hat \Gamma_\mu 
        &= \gamma^\mu \left(\Gamma_\mu + \frac{d-1}{2} \frac{\partial_\mu\Omega}{\Omega}\right).
\end{align}
The canonical fermion is defined as
\begin{align}
\psi_c := \Omega^{(1-d)/2} \psi.
\end{align}

Then we end up with the action in the canonical basis as
\begin{align}
	S 
        \supset \int d^dx\sqrt{-g}\left[ - i\bar\psi_c e_a^\mu \gamma^a D_\mu \psi_c \right].
\end{align}
Thus $\Omega$ factor cancels out even in $d$ dimensions, meaning that the massless fermion is conformal in any dimension. That is, the kinetic term of the fermion decouples from $\phi$.

\subsection{Vector}
For the gauge boson, the action is
\begin{align}
\label{appeq: gauge boson action}
	S 
        \supset \int d^dx\sqrt{-g}\frac{1}{\Omega^{d-4}}\left(-\frac{1}{4} g^{\mu\rho}g^{\nu\sigma} F_{\mu\sigma}^a F_{\nu\sigma}^a \right).
\end{align}
Thus the massless vector boson is conformal only in four dimensions.

\subsection{Yukawa}
The action of the Yukawa term becomes
\begin{align}
	S 
        &\supset \int d^dx\sqrt{-g}\frac{1}{\Omega^{d}}
            \left[y_{\psi} H \psi_{L} \psi_{R} + {\rm h.c.} \right], \nonumber \\
        &=\int d^dx\sqrt{-g} \,\Omega^{d-4 \over 2}
            \left[y_{\psi} H_c \psi_{Lc} \psi_{Rc} + {\rm h.c.} \right].
\end{align}
Thus, the $\Omega$ factor cancels out in the canonical basis in the four dimensions. Therefore, a massless fermion does not have any direct interaction with $\phi$ at the tree level.

\section{Interactions of secluded scalar}
\label{app: interactions of secluded scalar}

\subsection{Scalar-Higgs-Higgs interaction}
In the metric formalism, the relevant scalar-Higgs-Higgs interaction comes from terms Eqs.~\eqref{appeq: action in metric formalism} and \eqref{appeq: noncanonical Higgs action} in the non-canonical basis and \eqref{appeq: canonical Higgs action} in the canonical basis. After expanding $\Omega$ around $\phi=H=0$ in the flat spacetime, we obtain
\begin{align}
	S \supset \int d^4x \sqrt{-g} \frac{2\Lambda}{M_{\rm Pl}^2} \left(3\xi \Box\phi\,|H|^2 - \phi |\partial_\mu H|^2 \right).
\end{align}
in the non-canonical basis, 

In the canonical basis, this expansion yields
\begin{align}
\label{appeq: mixing between phi and canonical Higgs}
	S \supset \int d^4x \sqrt{-g} \frac{\Lambda}{M_{\rm Pl}^2} \left(6\xi -1\right) \Box\phi\,|H_c|^2.
\end{align}

In the Palatini formalism, the relevant interaction comes from only Eq.~\eqref{appeq: noncanonical Higgs action} in the non-canonical basis and \eqref{appeq: canonical Higgs action} in the canonical basis. Thus, after expanding $\Omega$ around $\phi=H=0$ in the flat spacetime,  we obtain
\begin{align}
S \supset 
    \int d^4 x  {- 2 \Lambda \over M_{\mr{pl}}^2} 
     \phi |\partial_{\mu} H|^2,
\end{align}
in the non-canonical basis and 
\begin{align}
S \supset 
    \int d^4 x {- \Lambda \over M_{\mr{pl}}^2} 
      (\Box \phi) |H_c|^2,
\end{align}
in the canonical basis.

\subsection{Scalar-Vector-Vector interaction}
To discuss the scalar-vector-vector interaction, we expand the $\Omega$ factor in the action of Eq.~\eqref{appeq: gauge boson action} with respect to $H$ and $\phi$ around $\phi = H = 0$ in the flat spacetime. This gives
\begin{align}
\label{appeq: gauge boson action}
S 
    \supset \int d^dx\, \sqrt{-g}\,
   &\frac{-\Lambda \phi}{M_{\mathrm{Pl}}^2}
     \left(- \tfrac{4-d}{4}\, g^{\mu\rho} g^{\nu\sigma} F^a_{\mu\nu} F^a_{\rho\sigma}\right),
\end{align}
and therefore, at tree level, this contribution vanishes in the limit $d \to 4$. This implies that $\phi$ does not couple directly to massless vector bosons at tree level.

However, a scalar–vector–vector interaction is induced at the loop level even for the massless vector bosons. This contribution can be classified into two categories. The first arises from the Higgs loop induced by the interaction~\eqref{appeq: mixing between phi and canonical Higgs}, as illustrated in Fig.~\ref{fig: higgs loop induced diagram}. The second originates from the running effect of the gauge coupling. The former vanishes in the limit $\xi \to 1/6$, whereas the latter is independent of $\xi$.\footnote{
We note that computing higher-order contributions to the scalar-vector-vector interaction corresponds to evaluating the gauge-field contribution to the trace of the energy–momentum tensor, $T^{\mu}_{\mu}$, at higher orders (so-called trace anomaly). In our model in the conformal limit, this contribution is determined solely by the running of the gauge coupling.
} 

\begin{figure}[h]
\centering
\begin{tikzpicture}
  \begin{feynman}
    \vertex (phi) {\(\phi\)};
    \vertex [right=2cm of phi] (v0);
    
    \vertex [above right=2cm of v0] (v1);
    \vertex [below right=2cm of v0] (v2);

    \vertex [right =1cm of v1] (A1) {\(A_\mu\)};
    \vertex [right =1cm of v2] (A2) {\(A_\nu\)};

    \diagram* {
      (phi) -- [scalar] (v0),
      (v1) -- [photon] (A1),
      (v2) -- [photon] (A2),

      (v0) -- [fermion] 
      (v2) -- [fermion] 
      (v1) -- [fermion] (v0),
    };
  \end{feynman}
  \end{tikzpicture}
  \begin{tikzpicture}
  \begin{feynman}
    \vertex (o);
    \vertex [above=1cm of o] (a);
    \vertex [below=1cm of o] (b);
    \vertex [left=2cm of b] (phi) {\(\phi\)}; 
    \vertex [right =1.5cm of a] (A1) {\(A_\mu\)};
    \vertex [right =1.5cm of b] (A2) {\(A_\nu\)};

    \diagram*{
      (phi) -- [scalar] (b),

      (a) -- [fermion, half left,  looseness=1.5] (b),
      (b) -- [fermion, half left, looseness=1.5] (a),
      (a) -- [photon] (A1),
      (b) -- [photon] (A2),
    };
  \end{feynman}
\end{tikzpicture}
  \begin{tikzpicture}
  \begin{feynman}
    \vertex (phi) {\(\phi\)};
    \vertex [right=1.5cm of phi] (a);
    \vertex [below right=1cm of a] (b);
    \vertex [below right=1cm of b] (c);
    \vertex [above right =1.5cm of a] (A1) {\(A_\mu\)};
    \vertex [below right =1cm of c] (A2) {\(A_\nu\)};

    \diagram*{
      (phi) -- [scalar] (a),
      (a) -- [photon] (b),

      (b) -- [fermion, half left,  looseness=1.5] (c),
      (c) -- [fermion, half left, looseness=1.5] (b),
      (a) -- [photon] (A1),
      (c) -- [photon] (A2),
    };
  \end{feynman}
\end{tikzpicture}
\caption{Anomaly-induced interaction with               fermion loop}
\label{fig: anomaly induced interaction from fermion loop}
\end{figure}
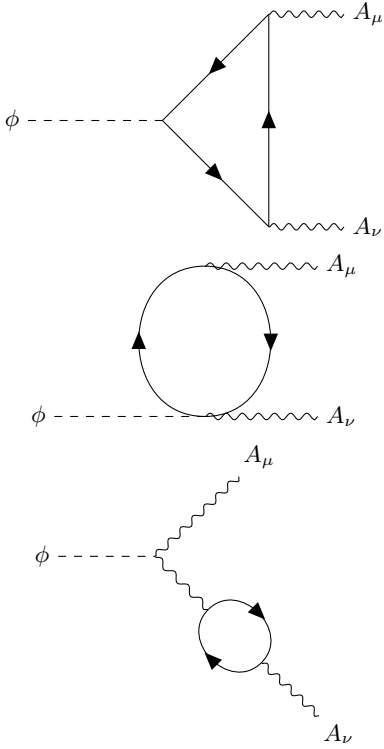

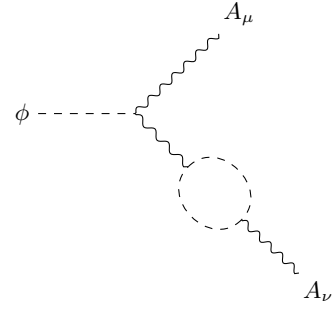
\begin{figure}[h]
\centering
  \begin{tikzpicture}
  \begin{feynman}
    \vertex (phi) {\(\phi\)};
    \vertex [right=1.5cm of phi] (a);
    \vertex [below right=1cm of a] (b);
    \vertex [below right=1cm of b] (c);
    \vertex [above right =1.5cm of a] (A1) {\(A_\mu\)};
    \vertex [below right =1cm of c] (A2) {\(A_\nu\)};

    \diagram*{
      (phi) -- [scalar] (a),
      (a) -- [photon] (b),

      (b) -- [scalar, half left,  looseness=1.5] (c),
      (c) -- [scalar, half left, looseness=1.5] (b),
      (a) -- [photon] (A1),
      (c) -- [photon] (A2),
    };
  \end{feynman}
\end{tikzpicture}
\caption{Anomaly-induced interaction with               scalar loop}
\label{fig: anomaly induced interaction from scalar loop}
\end{figure}

We now focus on the anomalous contribution that remains in the limit $\xi \to 1/6$, and restrict ourselves to the period before the electroweak phase transition. For the gauge boson associated with a $U(1)$ gauge symmetry, the relevant diagrams are shown in Figs.~\ref{fig: anomaly induced interaction from fermion loop} and ~\ref{fig: anomaly induced interaction from scalar loop}. Among the diagrams in Fig.~\ref{fig: anomaly induced interaction from fermion loop}, the first and second diagrams correspond to contributions that appear only in the non-canonical basis, and they originate from the interaction term in Eq.~\eqref{appeq: noncanonical fermion action}. The third diagram in Fig.~\ref{fig: anomaly induced interaction from fermion loop} and the diagram in Fig.~\ref{fig: anomaly induced interaction from scalar loop} are based on the interaction in Eq.~\eqref{appeq: gauge boson action}. 

As already mentioned, the interaction given in Eq.~\eqref{appeq: gauge boson action} vanishes at tree level in four dimensions. However, at the loop level, it produces a nontrivial contribution. The reason is that the loop diagram contains divergences, and dimensional regularization introduces a factor of $1/(2 - d/2)$.\footnote{
However, the Higgs-induced loop contribution shown in Fig.~\ref{fig: higgs loop induced diagram}, as well as the first two diagrams in Fig.~\ref{fig: anomaly induced interaction from fermion loop}, do not contain divergences.
}
This factor cancels the prefactor appearing in Eq.~\eqref{appeq: gauge boson action}, leaving a finite contribution. 

Alternatively, if one uses Pauli–Villars (PV) regularization, the scalar $\phi$ directly couples to the mass of the PV regulator field in the Einstein frame, thereby generating an effective scalar–vector–vector interaction.
In this picture, the $\phi$-dependence of the regulator mass contributes to the renormalization of the gauge coupling.

We note that the first two diagrams in Fig.~\ref{fig: anomaly induced interaction from fermion loop} cancel each other in the non-canonical basis. Consequently, only the third diagram induces the scalar–vector–vector interaction. In the canonical basis, $\phi$ does not have any direct interaction with fermions, and therefore only the third diagram contributes to the scalar–vector–vector interaction.

Then, the scalar–vector–vector interaction is given by~\cite{Endo:2007ih, Endo:2007sz, Watanabe:2010vy}
\begin{align}
\mathcal{L}_{\mathrm{eff}} \supset
\sqrt{-g}\, \frac{- \Lambda \phi}{M_{\mathrm{Pl}}^2} \Bigg(
   \frac{\beta(g')}{2 g'} B_{\mu\nu} B^{\mu\nu} 
 &+ \frac{\beta(g)}{2 g} W_{\mu\nu}^a W^{a\,\mu\nu} 
 \nonumber \\
 &+ \frac{\beta(g_s)}{2 g_s} G_{\mu\nu}^a G^{a\,\mu\nu} 
\Bigg),
\end{align}
where $\beta(g')$, $\beta(g)$ and $\beta(g_s)$ denote the beta function of each SM gauge interaction.

%%%%%%%%%%%%%%%%%%%%%%%%%%%%%%%%%%%%%%%%%%%%%%%%%%%%%%%%%%%%%%
%%%%%%%%%%%%%%%%%%%%%%%%%%%%%%%%%%%%%%%%%%%%%%%%%%%%%%%%%%%%%%

%%%%%%%%%%%%%%%%%%%%%%%%%%%%%%%%%%%%%%%%%%%%%%%%%%%%%%%%%%%%%%
%%%%%%%%%%%%%%%%%%%%%%%%%%%%%%%%%%%%%%%%%%%%%%%%%%%%%%%%%%%%%%

%%%%%%%%%%%%%%%%% Ref %%%%%%%%%%%%%%%%%%%%%%%%
%\bibliographystyle{utphysmod}
\bibliography{ref}
%%%%%%%%%%%%%%%%%%%%%%%%%%%%%%%%%%%%%%%%%%%%%%

\end{document}